*Full Length Research Paper*

# Particle model of single bubble sonoluminescence

## M. Adama Maiga

LML Laboratory (Laboratoire de Mécanique de Lille) / Arts et Metiers ParisTech, 8 Boulevard Louis XIV, 59046 Lille, France.

**The single bubble sonoluminescence is a phenomenon where the vapour or gas bubble trapped in a liquid collapse by emitting light. Sonoluminescence is most often characterized by the formation of an imploding shock and the plasma. However, studies show that in some cases neither an imploding shock nor a plasma has been observed. This study concerns a physical explanation of the Single Bubble Sonoluminescence (SBSL), where neither an imploding shock nor a plasma has been observed. A particles model of SBSL is proposed based on the dissociation hypothesis phenomenon and the experimental studies showed that particles can be trapped in bubble and to emit light during the evolution of a single bubble. During the bubble collapse phase, the bubble radius can reach a few nanometers. The bubble is considered a quantum dot, where the particles from the molecules dissociation, are trapped and confined in the bubble. This containment imposes on particles a strong excitement which allows them to gain a strong energy and to be in an excited state. Thus, at the moment of the bubble afterbounce, the particles de-excite by losing energy in the form of photon emission, which explains the light emission. The model shows that the energy of the particles is physically acceptable only if the bubble is spherical. The sphericity of the bubble, which is a necessary condition for the particles model of SBSL, is experimentally observed in the collapse phase (to the first afterbounce) but not in the afterbounces phase, where the bubble is unstable and not spherical. This explains why the bubble emits light to the collapse but not in the afterbounces phase where it is also very small. This condition of spherical bubble can constitute a validation of the particles model of SBSL. In this model, with the interaction between particles, it is also possible that the particles emit light by Bremsstrahlung radiation emission. But contrary to case where the temperature is very high, that is, where there is the formation of the plasma, in this case those are the size and sphericity of the bubble, that is, the containment, which impose on particles a strong excitement which allow them to move.**

**Key words**: Sonoluminescence, modeling, particles, quantum mechanics.

## INTRODUCTION

The Sonoluminescence is a phenomenon where the vapour or gas bubbles trapped in a liquid collapse by emitting light (photons). There are two types of Sonoluminescence: the Sonoluminescence emitted by

E-mail: Mahamadou.ADAMAMAIGA@ensam.eu.

multiple bubbles (Multiple Bubble Sonoluminescence, MBSL) (Frenzel and Schultes, 1934) and that emitted by a single bubble (Single Bubble Sonoluminescence, SBSL) highlighted by Gaitan et al. (1992). Bubbles are produced during a cavitation, phenomenon consists in the forming of vapour bubbles form in a fluid submitted to a local pressure drop (Peterson and Anderson, 1967; Levinsen, 2014; Su et al., 2003). This study focuses on the Single Bubble Sonoluminescence.

Single Bubble Sonoluminescence is a very complex phenomenon, which involves various physical mechanisms such as phase transition, heat transfers, chemical reactions, atoms ionization… Some studies helped to determine the experimental conditions to realize the SBSL. Thus, studies (Holt and Gaitan, 1996; Lohse et al., 1997; Gaitan and Holt, 1999) showed in a phase diagram a sonoluminescence stability area (the initial radius of the bubble as a function of the acoustic pressure amplitudes), where a bubble subjected to an acoustic pressure emits some light. In addition, the dissociation hypothesis that leads to stable SBSL in the water. This hypothesis based on considerations of chemical balance, emitted by Hilgenfeldt et al. (1996) and experimentally verified by many authors such as Matula and Crum (1998), Ketterling and Apfel (1998).

In physical point of view, studies of the SBSL phenomenon show that the extreme conditions of pressure and temperature which occur in the bubble during the collapse phase lead to:

$i$) the gas molecules dissociation (Nitrogen, Argon, Oxygen, etc.) present in the bubble. Also, experimental studies (McNamara III et al., 1999; Flannigan and Suslick, 2005; Flannigan and Suslick, 2007) highlight the ability of the particles atoms ($Ar, Ne, Xe, C_2$, etc.) and ions ($Ar^+, Xe^+, Kr^+$, etc.) to emit light during the sonoluminescence. Above all, these studies show that of the particles (electron, atoms, ions) can be trapped and especially emit light during the evolution of cavitation or the SBSL.

$ii$) on one hand the formation of an imploding shock (Wu and Roberts, 1993; Moss et al., 1999) and on the other hand the temperature which can reach thousands of Kelvins (McNamara III et al., 1999; Xu and Suslick, 2010; Levinsen, 2010), and that the light would be emitted by radiation of the ionized gas inside the bubble.

However, studies show that in some cases neither an imploding shock nor a plasma has been observed (Weninger et al., 1997; Barber et al., 1997; Wang et al., 1999; Putterman and Weninger, 2000), but especially even if the temperature can be high it is not enough to explain the spectrum observed.

As observed, a theoretical explanation of the SBSL phenomenon remains a subject of study. In a physical phenomenon study, the understanding of a particular case, most often allows a very real progress in the understanding of the general case. So, in this present study we are particularly interested in a physical explanation of SBSL, where neither an imploding shock nor a plasma has been observed. In order to achieve this objective, we consider $i$) the dissociation hypothesis phenomenon and $ii$) the experimental studies which show that particles can be trapped in bubble and to emit light during the evolution of a single bubble.

In the literature, the evolution of a single bubble is modelled by the Rayleigh-Plesset equation (Plesset, 1948). With the works of Van der Waals, it is known that the bubble is limited by a very thin interface, this last one can be a barrier for particles (atoms, ions, electrons…) contained in the bubble.

The idea developed in this study is that during the bubble collapse phase, depending on conditions, the bubble radius can reach a few nanometers. In these cases, the bubble size is of the same order of magnitude as the quantum dot one. Thus, the particles (atoms, ions, electrons), from the molecules dissociation, are trapped and confined within the bubble. With the uncertainty principle of Heisenberg ($\Delta x \Delta p_x \geq h$), this containment imposes on particles a strong excitement which allows them to gain a strong energy and to be in an excited state. So, at the bubble afterbounce moment, the particles de-excite by losing energy in the form of photon emission, which explains the light emission even though neither an imploding shock nor a plasma has been observed.

It is better to couple the evolution of the bubble and those of the particles (particles interaction, Schrödinger equation), but this remains a great challenge and requires to know at each moment the evolution of the potential of energy in the bubble. The objective of this study this is specially to see if the model can explain certain characteristics of the Single Bubble Sonoluminescence phenomenon. Thus, by assuming some hypotheses and with the quantum theory the energy of the particles can be determined.

The present paper is divided into two parts. First is analysis of the models for determining the energy of the particles followed by the results and discussions.

**MODELS**

The radial motion of a single bubble is governed by Rayleigh-Plesset equation (Plesset, 1948). With the same hypotheses that Putterman and Weninger[22], equation is defined as following:

$$R\ddot{R} + \frac{3}{2}\dot{R}^2 = \frac{1}{\rho}\left(p_g - p_0 - p_a\right) + \frac{R}{\rho c}\frac{d}{dt}\left(p_g - p_a\right) - \frac{4\mu \dot{R}}{\rho R} - \frac{2\sigma}{\rho R} \quad (1)$$

Where $R$ is the radius, $\dot{R}$ the normal velocity to the interface, $\sigma$ the surface tension, $\mu$ the dynamic viscosity, $\rho$ the density of liquid, $c$ the speed of sound, $p_0$ is the constant ambient pressure, $p_g$ the gas pressure and $p_a$ the ultrasound driving. $p_a$ modelled as a spatially homogeneous, standing sound wave:

$$p_a(t) = -p'_a \cos(\omega_a t), \quad (2)$$

Where $p'_a$ the acoustic pressure amplitudes and $\omega_a = \frac{2\pi}{T}$, with $T = \frac{1}{26.5} kHz$. Based on the initial conditions and the evolution of gas inside the bubble, the Equations (1) and (2) allow determining the evolution of the bubble.

There are many models on the modelization of the pressure and temperature of gas and the number of particles. Some works (Kwak and Na, 1996; Yasui, 1997; Yasui, 1999), in addition to the evolution of the pressure and temperature of gas, model the evolution of the number of particles and the formation of the plasma. It is specified that, in this present study, one is interested only to Single Bubble Sonoluminescence where neither an imploding shock nor a plasma has been observed. As Brenner et al. (2002) pointed out, from modeling simplifications are extremely useful tools in exploring the phase space of sonoluminescence. Simple models assuming a uniform bubble interior have been developed with increasing detail.

Therefore, for the modelization of the pressure and temperature of gas, we assume an adiabatic equation of state for the bubble interior (McNamara III et al., 1999; Barber et al., 1997; Brenner et al., 2002; Löfstedt et al., 1993):

$$p_g(t) = \left(p_0 + \frac{2\sigma}{R_0}\right)\left(\frac{R_0^3 - a^3}{R^3 - a^3}\right)^\gamma \quad (3)$$

$$T(t) = T_0 \left(\frac{R_0^3 - a^3}{R^3 - a^3}\right)^{\gamma-1} \quad (4)$$

Where $\gamma$ is the ratio of specific heats, $R_0$ is the initial bubble radius, $T_0$ is the ambient temperature and $a$ is the van der Waals hard-core radius determined by the excluded volume of the gas molecules.

Concerning the particles contained in the bubble, one is primarily interested on the energy that they can acquire. To define this energy, the following hypotheses are assumed:

i) The bubble remains spherical and its internal interface is assumed perfectly reflective and gives rise to shocks perfectly elastic;
ii) There are very few particles in the bubble, compared to the sonoluminescence case with the formation of the plasma, but especially no interaction between the particles and thus each particle can be considered as being alone;
iii) The particles move by making back-and-forth on an axis of bubble diameter length or describe a uniform circular and plan motion of radius equals to the bubble one.

With these hypotheses, the particles remains confined and well localized in the bubble, they are in a bound state. Thus, the energy of the particles can be considered to quantify. In addition, each particle can be considered as in a box. The motion and the energy of a particle in a box are well known, what is interesting this is the application in the Sonoluminescence phenomenon. If we assume the potential energy in the bubble equals zero, then the energy of a particle is given by that of a particle in an infinitely deep well or a particle in a one-dimensional ring.

**Particle energy in an infinitely deep well**

As the particle remains in the bubble, it can assume a particle with mass $m$ subjected to a potential energy $V(r)$ defined by:

$$\begin{cases} V(r) = 0 & \text{for} \quad -R < r < R \\ V(r) = +\infty & \text{for} \quad r < -R \text{ and } r > R \end{cases} \quad (5)$$

It is assumed that the particle makes back-and-forth with a constant velocity $v$, the particle motion is of one degree of freedom $r$ (Figure 1).

For more details on the theoretical study of the particle motion in the infinite potential well see Landau and Lifchitz (1967). The energy level of the particle is:

$$E_n = \frac{n^2 h^2}{32 m R^2} \quad (6)$$

Where $n$ is integer greater than or equal to one, $h$ Planck constant. The Equation (6) gives the energy of a particle which moves on an axis of length equal to the bubble diameter.

**Particle on a ring**

When the particle describes a uniform circular motion in a plan $Oxy$ (Figure 2) with $\varphi$ the azimuth, the energy level of the particle is given by the case of a particle on a one-dimensional ring (Flügge, 1971) (Equation 7):

$$E_n = \frac{n^2 h^2}{8\pi^2 m R^2} \quad (7)$$

The Equations (6) or (7) allow to determine the energy of a particle in an infinitely deep well or on a ring, the energy is a function of $1/R^2$.

The energy of the particle depends on the bubble radius. The radius variation allows the particle to emit on a wide range of wavelengths, from where the continuous spectra observed.

**RESULTS AND DISCUSSION**

This study would be positioned in the sonoluminescence stability area given by Gaitan and Holt (1999), where the intervals of the acoustic pressure amplitudes and the initial radius are $1.3\, atm \leq p'_a \leq 1.4\, atm$ and $2\, \mu m \leq R_0 \leq 7\, \mu m$, respectively.

Figure 3a presents for $p'_a = 1.35\, atm$ and $R_0 = 4.5\, \mu m$, dimensionless values of the evolutions of the acoustic pressure, the radius and the energy proportion $1/R^2$. For evolution of the bubble, the different phases of the bubble dynamics which is subjected to a sinusoidal pressure are well described by Putterman and Weninger (2000) and Brenner et al. (2002).

Figure 3b and c display the time evolution of the gas pressure with a zoom on the collapse moments of bubble and the one of the gas temperature, respectively. They show pressure and temperature peaks in the order of 30000 *bars* and 5500 *K* which occur over extremely short time periods.

Figure 3a-c show that in the growth phase of the bubble the pressure and temperature decrease and in its collapse phase the ones increase. The particles model of SBSL is consistent with the evolution of the pressure and temperature of the gas. Indeed, the pressure and temperature measure the particles collisions on the internal interface of the bubble and the degree of thermal motion (excitement) of the particles, respectively. Thus,

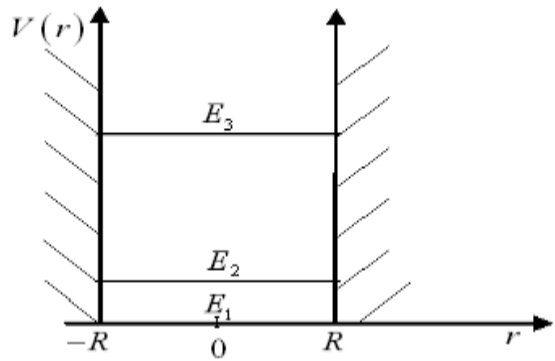

**Figure 1.** Back-and-forth motion on axis $r$. The hatched parts indicate forbidden zones for particle.

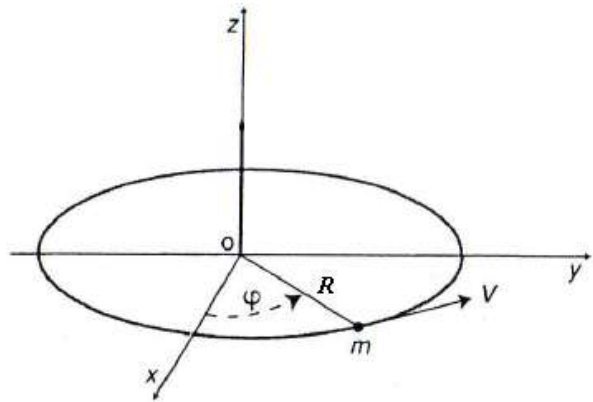

**Figure 2.** Uniform circular motion.

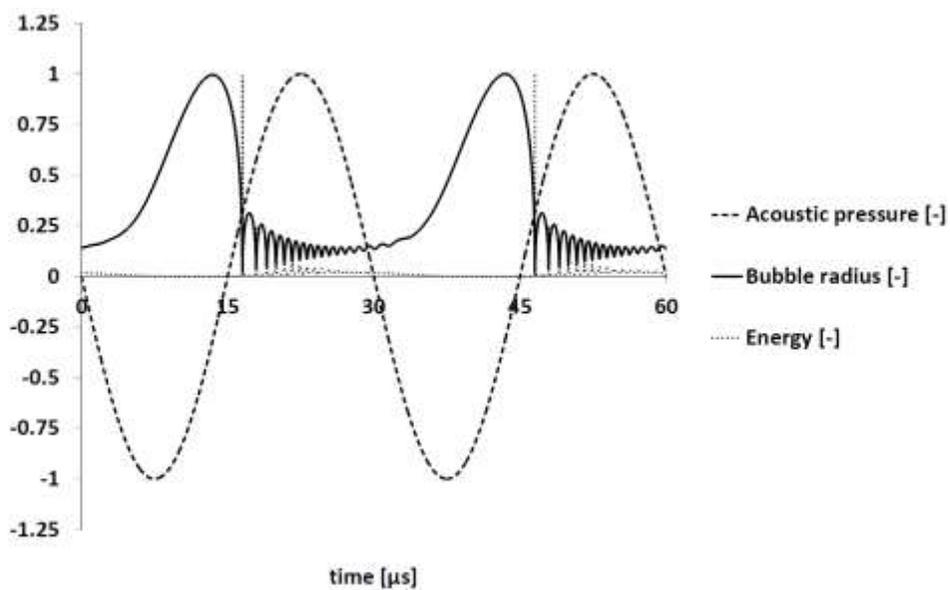

**Figure 3a.** Dimensionless values of the evolutions of the acoustic pressure, the radius and the energy for $p'_a = 1.35\ atm$, $R_0 = 4.5\ \mu m$, $R_0/a = 8.54$ and $T_0 = 300\ K$.

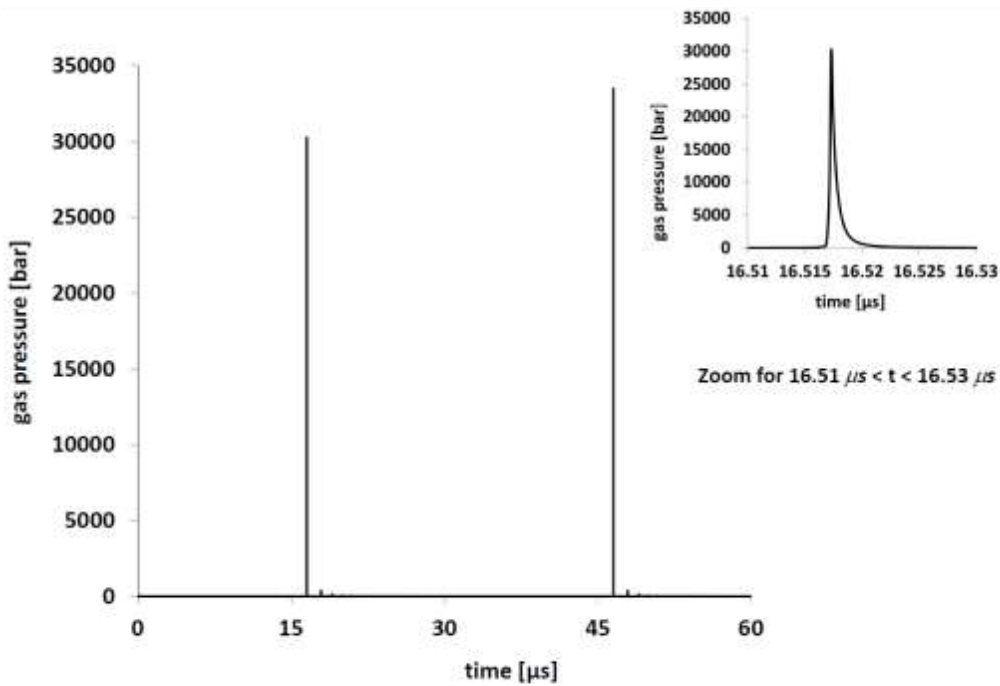

**Figure 3b.** Time evolution of the gas pressure inside the bubble for $p'_a = 1.35\ atm$, $R_0 = 4.5\ \mu m$, $R_0/a = 8.54$ and $T_0 = 300\ K$.

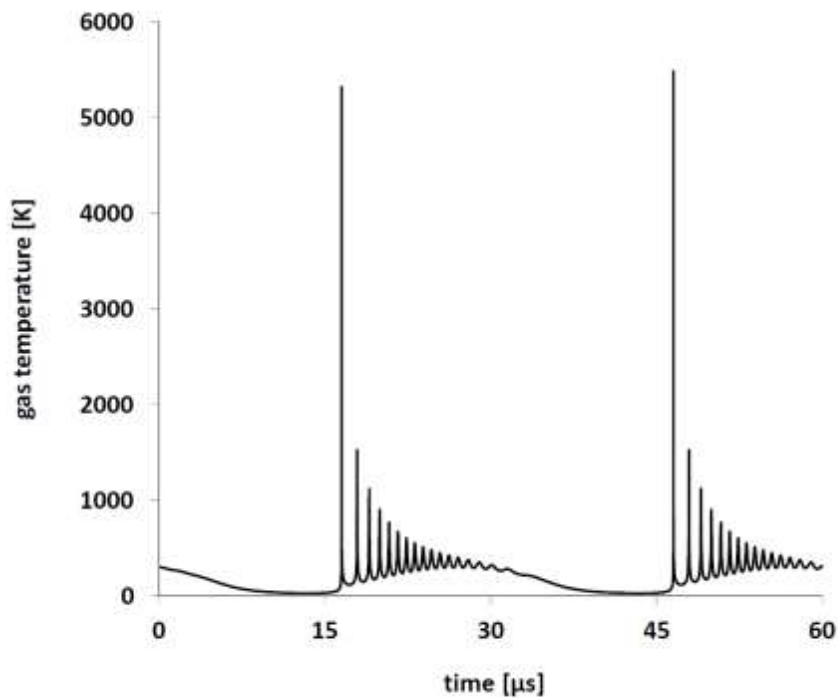

**Figure 3c.** Time evolution of the gas temperature inside the bubble for $p'_a = 1.35\ atm$, $R_0 = 4.5\ \mu m$, $R_0/a = 8.54$ and $T_0 = 300\ K$.

when the bubble grows the energy of the particles decreases, therefore the impact speeds and excitement of the particles that contributes to the evolution of the pressure and temperature decrease. When the bubble collapses, the energy of the particles increases, which will increase their impact speeds and degrees of excitement, thereby increasing pressure and temperature inside the bubble.

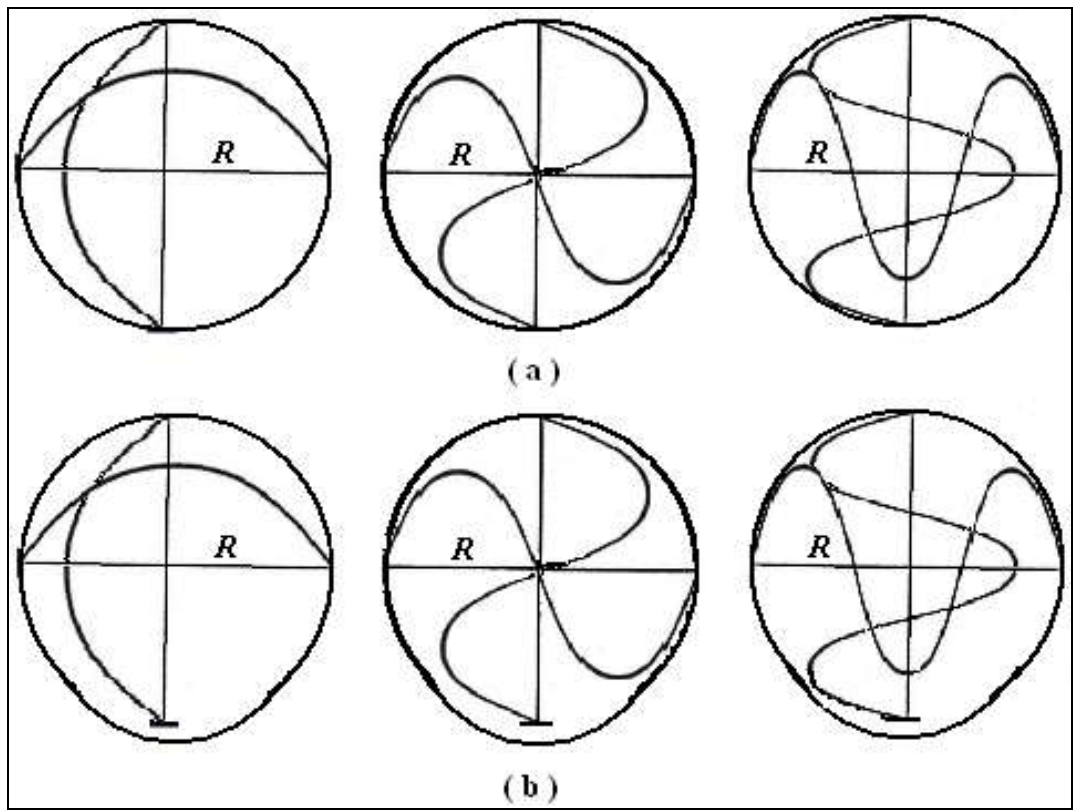

**Figure 4.** Wave of a vibrating string of length $2R$ inside a bubble: (a) acceptable physical state and (b) unacceptable physical state.

Concerning the energy $E_n$ of the particles, it can take all the proportional values to $1/R^2$ which are physically acceptable. As shown in Figure 3a, an energy peak to the first afterbounce of the bubble. This indicates exactly the moment when the bubble emits light. Thus, during the bubble collapse the particle gains a certain energy, which it loses to the first afterbounce by emitting light.

In both motions cases, back-and-forth and circular, the particle is confined in the bubble as a particle in the box. As the wave theory of sound vibrations or electromagnetic field inside a cavity, which is a problem giving rise to standing waves, in the bubble case a physically acceptable condition would also lead to standing waves.

To have standing waves, it is necessary for:

$i$) the back-and-forth motion case, that $2R$ is equal to $n_w \lambda/2$: $2R = n_w \lambda/2$, with $n_w$ integer number. Thus, for that a particle has a unique wavelength, whatever the direction of motion, the bubble must be perfectly spherical. For a vibrating string of length $2R$ inside a bubble, Figure 4 represents: (a) acceptable physical state and (b) unacceptable physical state.

$ii$) the circular motion case, that the ratio between the circumference of the ring and wavelengths must be equal to an integer: $2\pi R = n_w \lambda$. Figure 5 shows the wave of a particle on a ring: (a) acceptable physical state and (b) unacceptable physical state.

To obtain a physically acceptable condition, the wavelength of the particle must be unique. Whatever the direction of particle motion in the bubble, the theoretical analysis shows that the bubble should remain stable and especially spherical. In single bubble sonoluminescence, studies (Gaitan and Holt, 1999; Hilgenfeldt et al., 1996; Barber et al., 1997; Brenner et al., 2002; Matula, 1999) show that the bubble is stable and spherical during the collapse phase (to the first afterbounce). However, Gaitan and Holt (1999) show that the bubble is unstable and not spherical during the afterbounces phase. This study can explain why in the afterbounces phase, where the bubble can be the same size as to the first afterbounce, is not observed the emission of light.

One notes that even if the motion path of the particle is an ellipse inscribed in a circle of bubble radius, the path (perimeter) which depends of the bubble radius can be equal to an integer of the wavelengths. To do this, the bubble sphericity condition is necessary to obtain a unique wavelength (a physically acceptable situation) whatever the direction of particle motion in the bubble.

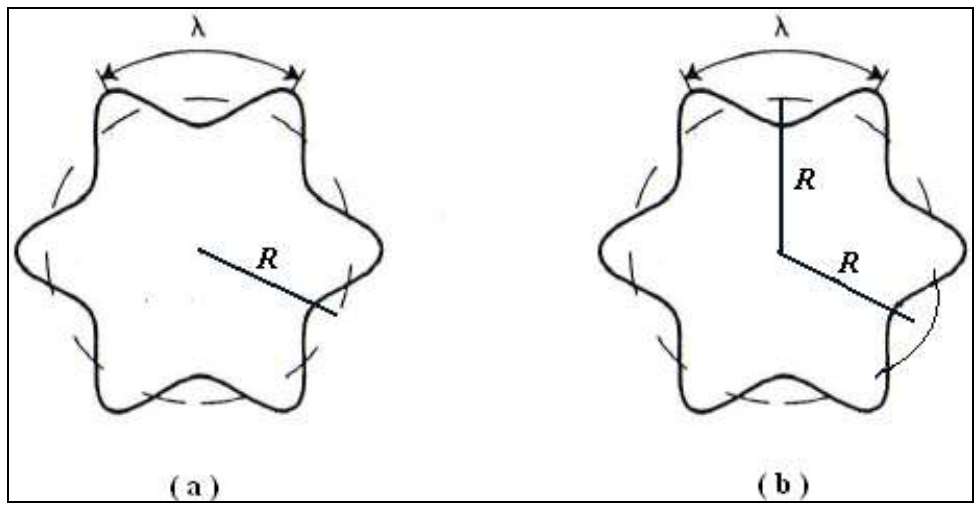

**Figure 5.** Wave of a particle on a ring: (a) acceptable physical state and (b) unacceptable physical state.

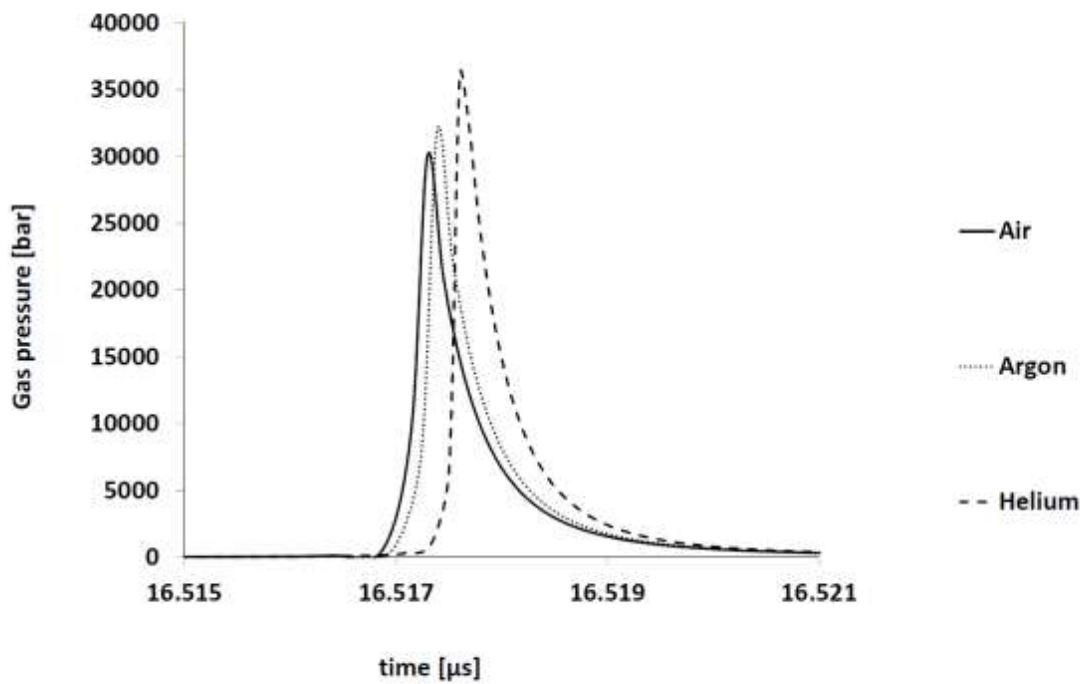

**Figure 6.** Time evolution of the gas pressure inside the bubble for different gas with $p'_a = 1.35\ atm$, $R_0 = 4.5\ \mu m$, and $T_0 = 300K$: ($R_0/a = 8.54$ for air, $R_0/a = 8.86$ for argon, $R_0/a = 10$ for helium).

An investigation of the influence of the main parameters of a sound sonoluminescence bubble was explored. Figures 6 and 7 display a zoom on the time evolution of the pressure and temperature of the gas for different $R_0/a$ (that is,. For different gas inside the bubble), respectively. The results show that the more $R_0/a$ increases the more the gas pressure is high. However, they show that the gas temperature was very little influenced by the $R_0/a$.

For the application of the particles model of Single Bubble Sonoluminescence, a free electron particle of mass $m = 9.1094\ 10^{-31}\ kg$ is considered. Concerning the energy level $n$, it may be high because a particle as a free electron passes of a low energy level (where it is bound to the nucleus) to almost an infinite level (where it is free with initial energy). For this application, as in experimental studies where neither an imploding shock nor a plasma has been observed, it is considered the variation of temperature low thus $\gamma = 1$.

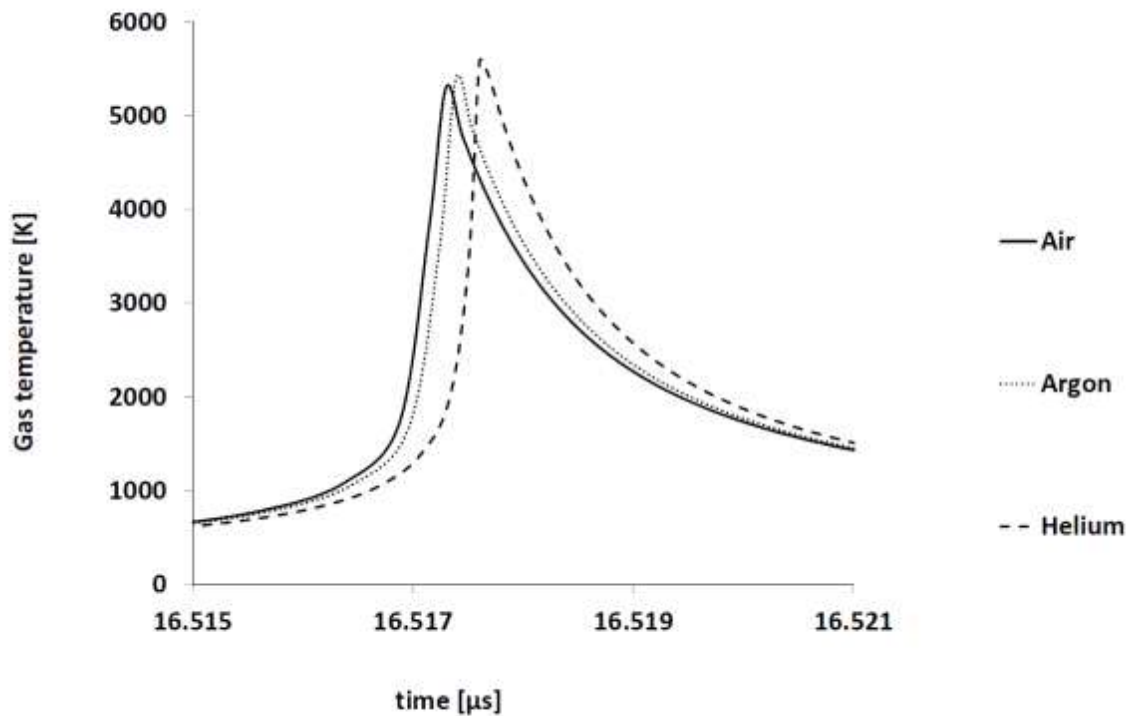

**Figure 7.** Time evolution of the gas temperature inside the bubble for different gas with $p'_a = 1.35\ atm$, $R_0 = 4.5\ \mu m$, and $T_0 = 300\ K$: ($R_0/a = 8.54$ for air, $R_0/a = 8.86$ for argon, $R_0/a = 10$ for helium).

Studies (Hiller et al., 1992; Didenko and Suslick, 2002) show that the energy of the emitted photon is around 6 eV. With a bubble of $R_0 = 4.5\ \mu m$ which is subjected to an acoustic pressure of $p'_a = 1.35\ atm$, the simulation gives a minimum radius of $R_{min} \simeq 34\ nm$. This minimum radius value is of the same order of magnitude as the size of certain quantum dots studied by Ngo et al. (2006) and Movilla et al. (2009). The minimum radius leads to a $n$ in the order of 272 and 427 for back-and-forth and circular motions, respectively. The hypothesis to determine the energy level $n$ remains strong. This hypothesis allows, above all, to estimate the energy of the particle and to study the influence of the acoustic pressure amplitudes $p'_a$.

With these energy levels, for $R_0 = 4.5\ \mu m$ and by varying $p'_a$, Figure 8 shows that the more $p'_a$ is high the more minimum radius is low and the energy $E_n$ is high.

Figure 9 presents a zoom on the evolution of $E_n$, for $p'_a = 1.35\ atm$ and $R_0 = 4.5\ \mu m$. It shows that the time interval between $6\ eV$ and $2\ eV$ energy (which can be considered as the light emission duration of the bubble) is in the order of $48\ ps$. This is in the same order that the shortest pulses measure by Gompf et al. (1997) and Hiller et al. (1998) which is of $60\ ps$ ($\pm 6\ ps$) and $40\ ps$, respectively. Also, Gompf et al. (1997) assert that one cannot exclude that there is a stable regime SL where the pulse width is shorter than $50\ ps$. Before, Barber and Putterman (1991) by using faster detectors show that the width of the sonoluminescence spike is in fact less than $50\ ps$.

**Conclusions**

This work is dedicated to the theoretical study of Single Bubble Sonoluminescence (SBSL), where neither an imploding shock nor a plasma has been observed. New model of the particles model of Single Bubble Sonoluminescence, is proposed. The model is based on the dissociation hypothesis phenomenon and the experimental studies which show that particles can be trapped in bubble and to emit light during the evolution of a single bubble.

Thus, during the evolution of the bubble some particles, from the molecules dissociation, are trapped and confined in the bubble. During the collapse phase, the bubble size can be of the same order of magnitude as the quantum dot one. This containment imposes on particles a strong excitement which allows them to gain a strong energy and to be in an excited state. So, at the bubble afterbounce moment, the particles de-excite by losing energy in the form of photon emission. This behaviour explains the light emission even though neither an imploding shock nor a plasma has been observed.

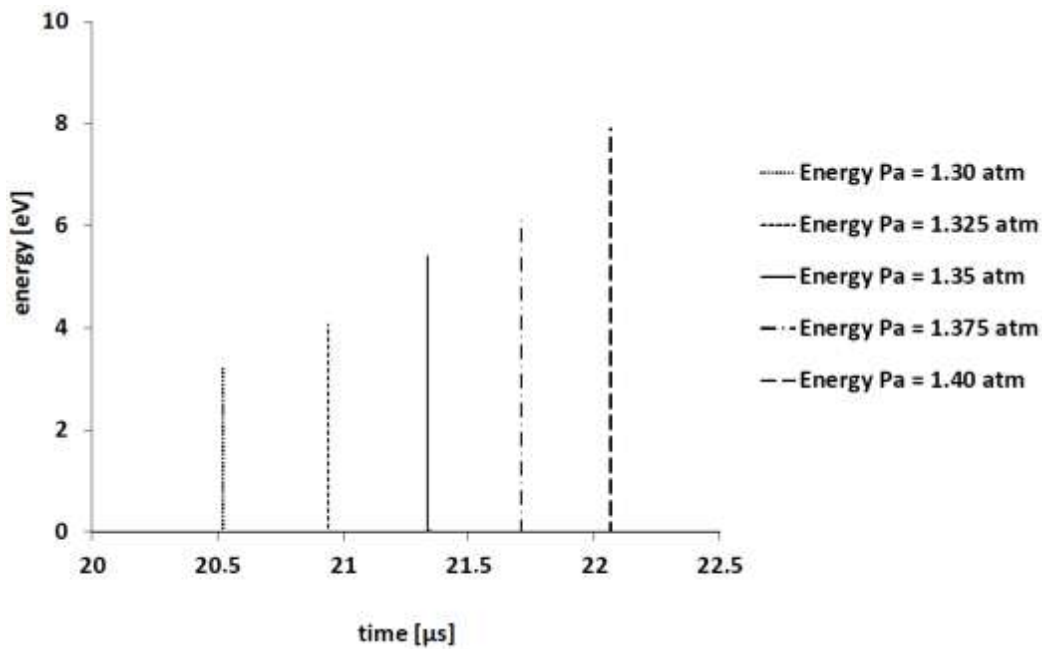

**Figure 8.** Evolution of the energy in the bubble collapse zone for $R_0 = 4.5\ \mu m$ and different $p'_a$.

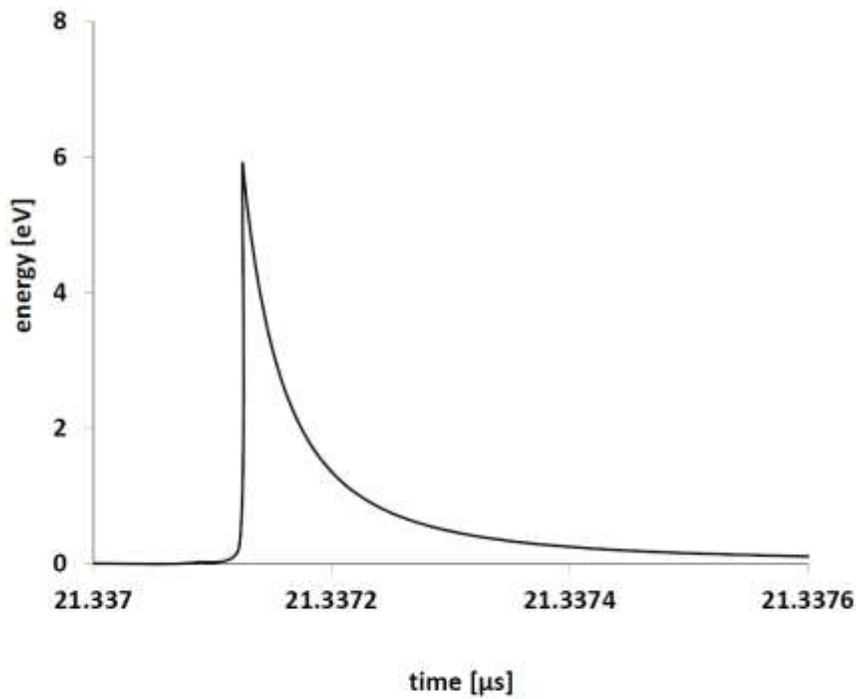

**Figure 9.** Zoom evolution of the energy in the bubble collapse zone for $p'_a = 1.35\ atm$ and $R_0 = 4.5\ \mu m$.

With some hypotheses and the quantum theory the energy of the particles was determined. This energy is physically acceptable that if the bubble is spherical. The sphericity of the bubble, which is a necessary condition for the particles model of Single Bubble Sonoluminescence, is experimentally observed in the collapse phase (to the first afterbounce) but not in the afterbounces phase, where the bubble is unstable and not spherical. This can explains why the bubble emits light to the collapse but not in the afterbounces phase where it is also very small. This condition of spherical bubble can constitute a validation of the particles model of Single Bubble Sonoluminescence.

The energy of the particles depends of the bubble radius. The radius variation allows the particles to emit on a wide range of wavelengths, from where the continuous spectra experimentally observed.

Study shows that the duration of the light emission which is estimated with the particles model of SBSL is of the same order as the picosecond flashes with is experimentally observed.

It can especially conclude that, if the evolution of the potential energy $V(r)$ in the bubble is known, the energy of the particle can be determined with the Schrödinger equation.

With the interaction between particles, it is also possible that the particles emit light by Bremsstrahlung radiation emission. But contrary to case where the temperature is very high, that is, where there is the formation of the plasma, in this particular case those are the size and sphericity of the bubble, that is, the containment, which impose on particles a strong excitement which allow them to move.

## Conflict of Interests

The author has not declared any conflict of interests.